\documentclass{aa}  

\usepackage{natbib}
\usepackage{graphicx}
\usepackage{txfonts}
\newcommand{\be}{\begin{equation}}
\newcommand{\ee}{\end{equation}}
\newcommand{\ba}{\begin{eqnarray}}
\newcommand{\ea}{\end{eqnarray}}

\begin{document}


\title{Time dependence of Fe/O ratio within a 3D Solar Energetic Particle propagation model including drift}

\titlerunning{Drift and time dependence of Fe/O ratio}

   \author{S.~Dalla\inst{1}
          \and
           M.S.~Marsh\inst{2}
          \and
           P.~Zelina\inst{1}
          \and 
           T.~Laitinen\inst{1}
          }

   \institute{Jeremiah Horrocks Institute, University of Central Lancashire,
    Preston, PR1 2HE, UK \\
              \email{sdalla@uclan.ac.uk}
         \and
            Met Office, Exeter, EX1 3PB, UK \\            
             }

   \date{Received March 2016; }

 
  \abstract
   {The intensity profiles of iron and oxygen in Solar Energetic Particle (SEP) events often display differences that result in a decreasing Fe/O ratio over time. The physical mechanisms behind this behaviour are not fully understood, but these observational signatures provide important tests of physical modelling efforts.}
   {In this paper we study the propagation of iron and oxygen SEP ions using a 3D model of propagation which includes the effect of guiding centre drift in a Parker spiral magnetic field. We derive time intensity profiles for a variety of observer locations and study the temporal evolution of the Fe/O ratio.}
   {We use a 3D full orbit test particle model which includes scattering. The configuration of the interplanetary magnetic field is a unipolar Parker spiral. Particles are released instantaneously from a compact region at 2 solar radii and allowed to propagate in 3D.
    } 
   {Both Fe and O experience significant transport across the magnetic field due to gradient and curvature drifts. We find that Fe ions drift more than O ions due to their larger mass-to-charge ratio, so that an observer that is not magnetically well connected to the source region will observe Fe arriving before O, for particles within the same range in energy per nucleon. As a result, for the majority of observer locations, the Fe/O ratio displays a decrease in time.
   }
   {We conclude that propagation effects associated with drifts produce a decay over time of the Fe/O ratio, qualitatively reproducing that observed in SEP event profiles.}


   \maketitle
%
%

\section{Introduction}

While protons and electrons are the main species in Solar Energetic Particle (SEP) events, often detected in the interplanetary medium following flares and Coronal Mass Ejections (CMEs), ions with mass number $A$$>$1 are also present.
Heavy ion observations display a wealth of signatures that can be used to infer the properties of the acceleration and propagation processes acting on the particles before they reach an observer at 1 AU.

One of the observational signatures that is often emphasized is the temporal evolution of the Fe/O ratio: this ratio has been shown, by many observers, to decay over time over the duration of an SEP event \citep{Sch1978,Mas2006,Mas2012,Zel2015}.  For an example of typical time profiles of Fe and O intensities and Fe/O ratio, see Figures 1 and 2 of \cite{Mas2006}.
In many so-called gradual events, thought to be associated with acceleration at CME-driven interplanetary shocks, the Fe/O ratio at the beginning of the event can have values considered typical of flare-associated events, and decay over time to values near or below the typical average value for gradual events \citep{Tyl2013}.

Several different explanations for the observed Fe/O decay over time have been put forward. 
In the first reports of the effect, it was proposed that it results from the rigidity dependence of the scattering mean free path $\lambda$ \citep{Sch1978}, whereby  $\lambda_{Fe}$$>$$\lambda_O$ due to the larger mass-to-charge ratio, $m$/$q$, of Fe in SEP events. In this interpretation, ions are assumed to be tied to the magnetic field line onto which they are injected:  as a result a single spatial variable, the distance travelled along the magnetic field line, is thought to be sufficient to describe propagation, so that the modelling is spatially 1D. A larger mean free path means that Fe ions arrive first at the spacecraft, so that their number is enhanced at the start of the event. 
A recent 1D focussed transport model including a mean free path proportional to  ($m$/$q$)$^{1/3}$ was shown to reproduce the observed heavy ion ratio time variations in several SEP events \citep{Mas2012}.

Other authors have suggested that the temporal characteristics of Fe/O are a signature of the acceleration process.  \cite{Can2003} proposed that the high Fe/O ratio values at the beginning of an SEP event are due to a flare-accelerated SEP component, while later lower values are due to SEPs accelerated by the CME-driven shock. A model of acceleration at a CME-driven shock in the presence of self-generated waves produced heavy ion profiles with a variety of temporal behaviours, similar to those observed in the 20 April 1998 event \citep{Ng1999}.

   \begin{figure*}
   \centering
   \includegraphics[scale=.67]{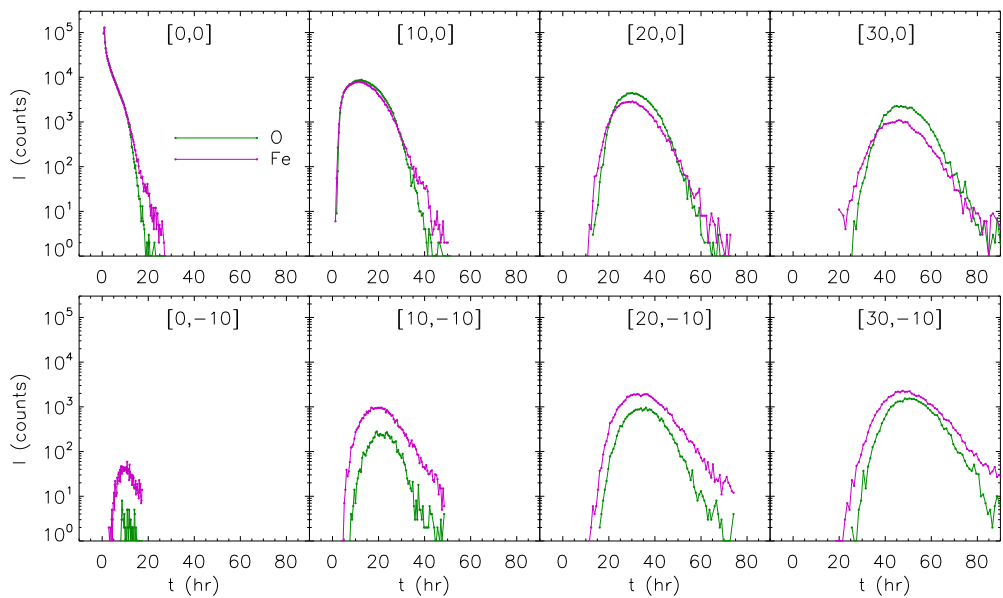}
   \caption{Fe ({\it magenta line}) and O ({\it green line}) intensities versus time for the energy range 10--30 MeV nucleon$^{-1}$ at various 1 AU locations relative to the magnetic field line connected to the centre of the injection region at the Sun. Labels in each panel give the observer's angular position as $[\Delta\phi_{1AU}, \Delta\delta_{1AU}]$, where $\Delta\phi_{1AU}$ is the 
heliographic longitude and $\Delta\delta_{1AU}$ the heliographic latitude relative to the
position of the Parker spiral field line through to the centre of the injection region. Here $\lambda$=1 AU. The same number of Fe and O ions were injected.}
\label{fig.intens1}  
     \end{figure*}

   \begin{figure*}
   \centering
   \includegraphics[scale=.7]{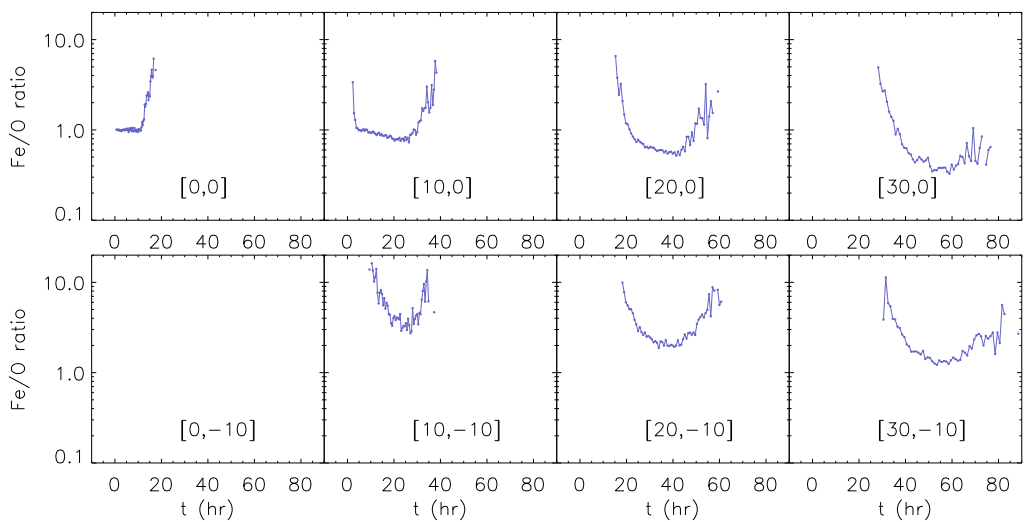}
   \caption{Fe/O ratio versus time for the energy range 10--30 MeV nucleon$^{-1}$ for the same 1 AU locations and parameters as in Figure \ref{fig.intens1}. The Fe/O ratio is calculated only for time intervals during which at least 10 Fe ions and 10 O ions were detected.}
\label{fig.ratio1}  
     \end{figure*}
%

   \begin{figure}
   \centering
   \includegraphics[scale=.7]{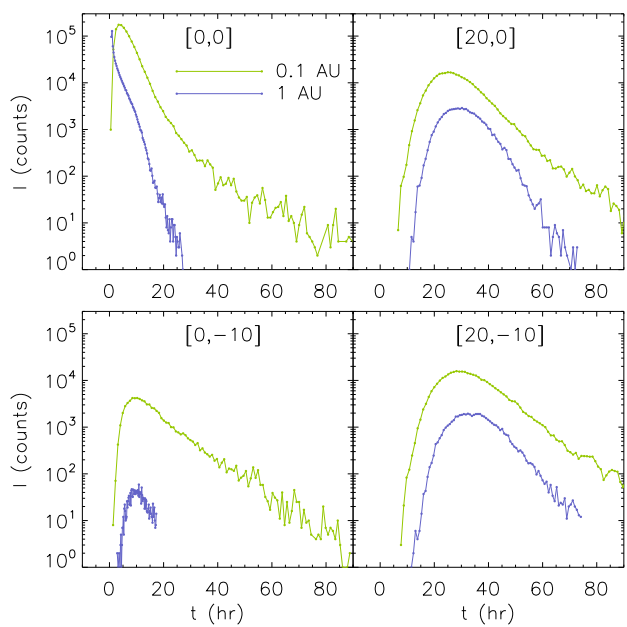}
   \caption{Fe intensities versus time for $\lambda$=0.1 AU ({\it green}) and $\lambda$=1 AU ({\it blue}), for the energy range 10--30 MeV nucleon$^{-1}$ for four representative 1 AU observer locations.}
\label{fig.varylambda}  
     \end{figure}
%

   \begin{figure*}
   \centering
   \includegraphics[scale=.7]{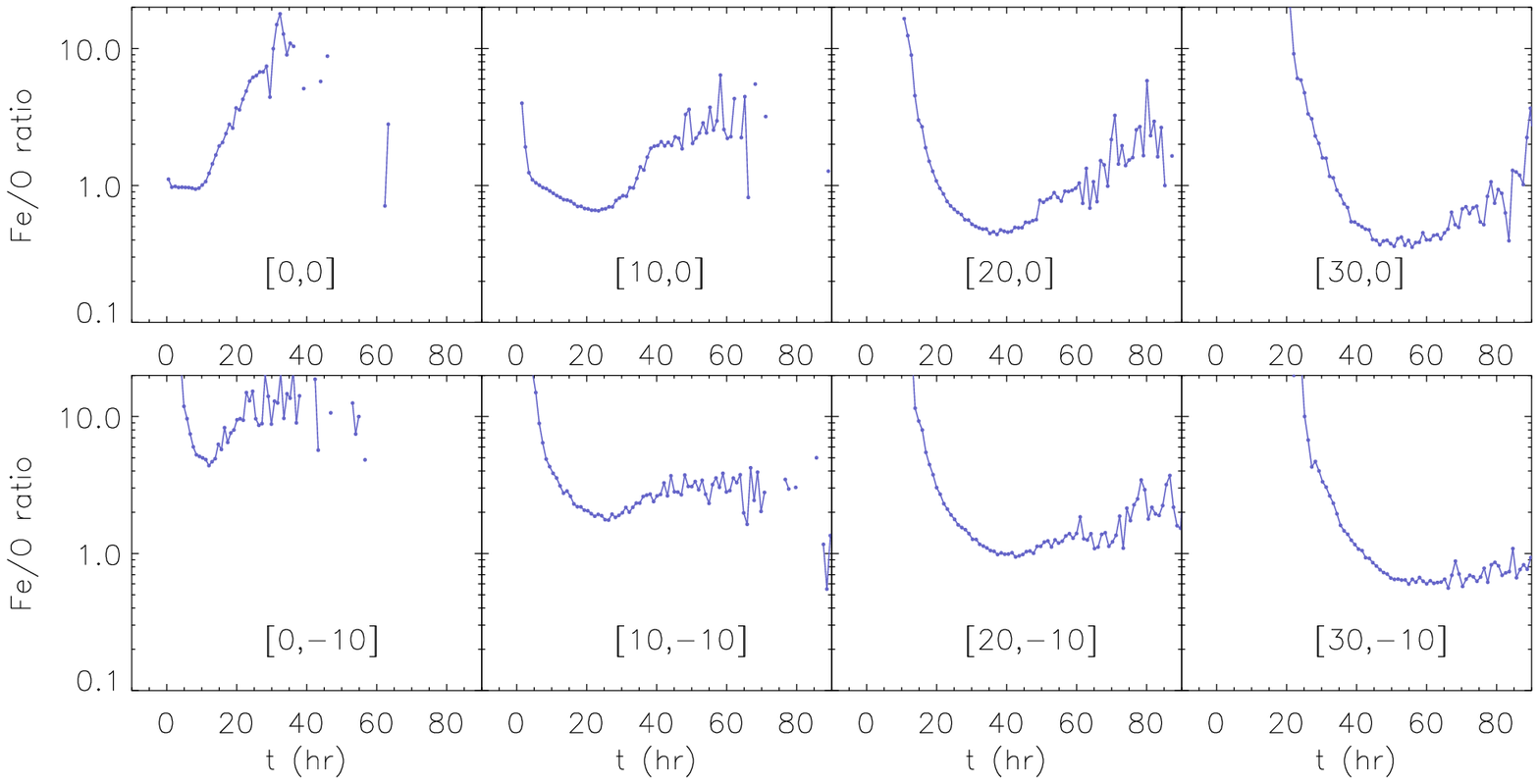}
   \caption{Fe/O ratio versus time for the energy range 10--30 MeV nucleon$^{-1}$ for the same 1 AU parameters as in Figure \ref{fig.ratio1} and mean free path $\lambda$=0.1 AU.}
\label{fig.ratio_lambda0.1}  
     \end{figure*}

\cite{Mas2006} showed that when Fe intensity profiles at 273 keV nucleon$^{-1}$ and 12 MeV nucleon$^{-1}$ are compared with O profiles in channels with average energy per nucleon double that of Fe, the differences between Fe and O profiles, and therefore the decay in time of the Fe/O ratio, disappeared.
They concluded that Fe/O decays are a result of interplanetary transport effects and cannot be explained by rigidity dependent acceleration and release from the source region.
 
An analysis of SEP measurements at Ulysses and Wind \citep{Tyl2013} showed that a qualitatively similar decrease in the Fe/O ratio can be detected at widely separated locations in interplanetary space. The authors concluded that the observed behaviour is the result of propagation effects. 

In this paper, we study the propagation of partially ionised iron and oxygen SEP ions in a simplified model of the interplanetary magnetic field, by solving their trajectories by means of a test particle code. Here all 3 spatial variables are retained in the description, making the modelling spatially 3D. Therefore, unlike traditional 1D models, we allow for the possibility that particles might leave the field line on which they were initially injected.
Our previous work has shown that, within a 3D model, particles experience transport perpendicular to the magnetic field due to drifts associated with the gradient and curvature of the Parker spiral magnetic field \citep{Mar2013, Dal2013}. Partially ionised heavy ions drift significantly more than protons (at the same energy per nucleon) due to their larger $m$/$q$. Here, for the first time, we derive the 3D propagation of Fe and O populations and study the Fe/O ratio at 1 AU as measured by several observers at locations of varying magnetic connection to the injection region. Some initial results related to this work were presented by \cite{Dal2015_icrc}.



\section{Simulations}\label{sec.test_simul}

Simulations are carried out by means of a 3D full-orbit test particle code, which integrates charged particle trajectories through a unipolar (outward pointing) Parker spiral interplanetary magnetic field (IMF) \citep{Mar2013}. Injection is instantaneous and from a compact region of angular extent 6$^{\circ}$$\times$6$^{\circ}$, located at $r$=2 $R_{sun}$. The heliographic longitude and latitude of the center of the injection region are $\phi$=0$^{\circ}$ and $\delta$=20$^{\circ}$ respectively.

We inject the same number $N$=$10^{6}$ of Fe and O ions. The measured charge states of Fe SEPs can vary within a rather wide range of values, depending on the event. In the simulations presented here we chose a charge state for iron $Q_{Fe}$=15, while for oxygen $Q_{O}$=7, consistent with typical SEP measured charge states \citep{Luh1985}. This gives  mass-to-charge ratios $(A/Q)_{Fe}$=3.7 and $(A/Q)_{O}$=2.3.
The injection spectrum of the heavy ions has a power law shape in energy per nucleon, with spectral index $\gamma$=1.1, in the range 10--400 MeV nucleon$^{-1}$. Other parameters of the runs are the same as in \cite{Mar2013}.

Within our simulations, a low level of scattering is introduced, with  a mean free path $\lambda$=1 AU. The value of  $\lambda$ is the same for the different species, i.e.~any rigidity dependence of the mean free path is neglected in our simulation. We do this deliberately to isolate the effects of drifts from those that would be caused by a rigidity dependent mean free path. No scattering across the magnetic field is present in our model.

Fe and O ion trajectories are integrated up to a final time $t_f$=100 hrs.
Drifts due to the gradient and curvature of the Parker spiral magnetic field cause a significant fraction of particles to propagate outside the flux tube delimited by the corners of the injection region, experiencing transport perpendicular to the magnetic field \citep{Mar2013,Dal2015}.
Because of their larger mass-to-charge ratio ($(A/Q)_{Fe}$=1.6 $(A/Q)_{O}$), and consequently larger drift velocity at the same energy per nucleon, Fe ions move across the field more efficiently than O ions.
Scatter plots showing the locations of Fe and O ions at the final time were presented by \cite{Dal2015_icrc}.

Figure \ref{fig.intens1} shows profiles of Fe and O counts versus time for the energy range 10--30 MeV nucleon$^{-1}$, for several 1 AU observers. 
Labels in each panel specify the observer's angular location at 1 AU from the Sun as $[\Delta\phi_{1AU}, \Delta\delta_{1AU}]$, where $\Delta\phi_{1AU}$ is the 
heliographic longitude and $\Delta\delta_{1AU}$ the heliographic latitude relative to the
position of the Parker spiral field line through to the centre of the injection region at the Sun. Therefore [0,0] corresponds to an observer's location directly connected to the centre of the injection region, and the other panels to less well connected observers. Panels to the right of [0,0] correspond to observers at the same latitude and more Western longitudes (i.e.~the source region is more Eastern relative to the observer), and panels below it show observers at latitudes further south.
The collecting area for each profile is 10$^{\circ}$$\times$10$^{\circ}$.

Figure \ref{fig.intens1} shows that significant heavy ion intensities are detected by observers that are not directly connected to the injection region (see all panels apart from [0,0]). Hence heavy ion propagation is taking place in 3D and not only along magnetic field lines, as is conventionally assumed. 
At the majority of not well connected observers Fe arrives earlier than O, because of its larger drift velocity at the same energy per nucleon. Fe also tends to peak earlier.
One can see that moving from left to right in the top row of Figure \ref{fig.intens1}, peak intensities tend to decrease, while they increase going from left to right in the bottom row, corresponding to latitudes below that of the injection region. This behaviour results from drift in latitude which is downwards for the unipolar outward-pointing magnetic field used here \citep{Mar2013, Dal2013}. 

Figure \ref{fig.ratio1} shows the time evolution of the Fe/O ratio for the same locations and energy ranges as in Figure \ref{fig.intens1}.
Here we can see that, at the majority of locations, the Fe/O ratio displays a decrease over time early in the event. 
It should be noted that since the same number of Fe and O ions are followed in our simulation, the  injection Fe/O ratio is 1.
Towards the end of the event, in many cases the Fe/O ratio displays an increase over time. This is due to the fact that the overall longitudinal extent of the flux tubes filled with O ions is smaller than for the case of Fe, due to smaller drift, resulting in a faster decay of O compared with Fe.

   \begin{figure*}
   \centering
   \includegraphics[scale=.67]{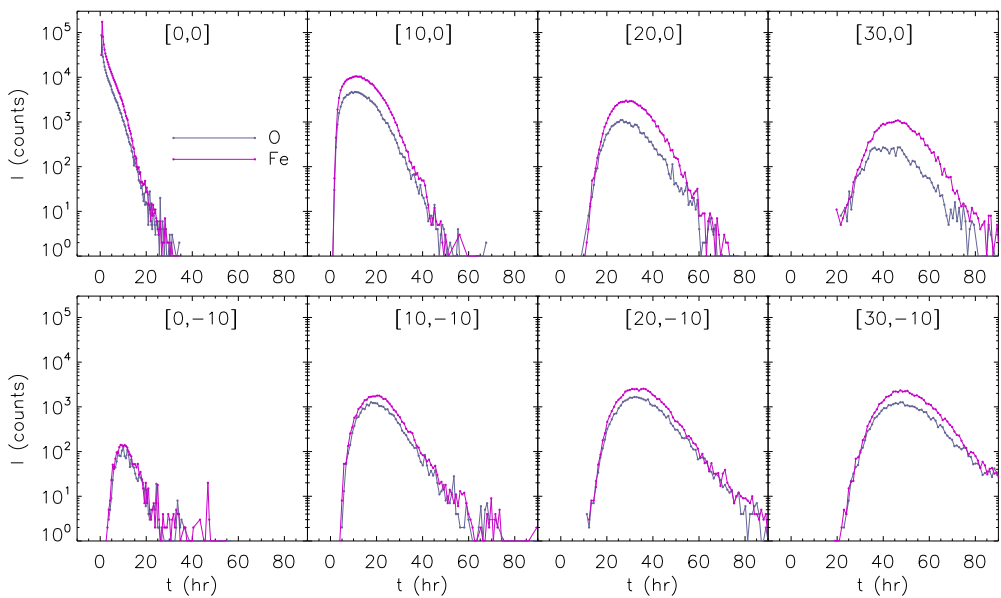}
   \caption{SEP intensities versus time for Fe in the energy range 10--30 MeV nucleon$^{-1}$ ({\it magenta line}) and O in the range 30--50 MeV nucleon$^{-1}$ ({\it blue line}) for the same observer locations as in Figure \ref{fig.intens1}.}
\label{fig.intens1_double_energy}  
     \end{figure*}

   \begin{figure*}
   \centering
   \includegraphics[scale=.7]{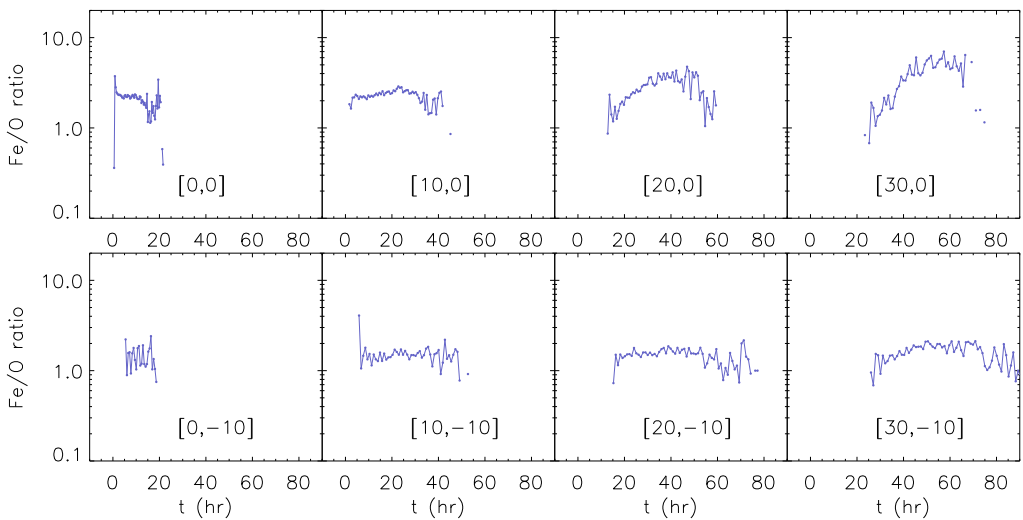}
   \caption{Fe/O ratio versus time obtained by considering Fe in the energy range 10--30 MeV nucleon$^{-1}$ and O in the range 30--50 MeV nucleon$^{-1}$, as in Figure \ref{fig.intens1_double_energy}.}
\label{fig.ratio1_double_energy}  
     \end{figure*}

An important question is whether the overall qualitative behaviour shown in Figures \ref{fig.intens1} and  \ref{fig.ratio1} for a mean free path $\lambda$=1 AU will change for different scattering conditions.
\cite{Mar2013} analysed proton drift across the magnetic field for $\lambda$=0.3, 1 and 10 AU, and showed that drift behaviour is very similar in the three situations and therefore only weakly dependent on the scattering conditions.
To study the effect of a different choice of mean free path on our results, we performed simulations of Fe and O propagation for $\lambda$=0.1 AU for the same ionic parameters considered earlier. 
Figure \ref{fig.varylambda} shows the effect of varying the mean free path on the 1 AU intensity profiles of Fe ions only, for four representative observer locations, where the green lines are for $\lambda$=0.1 AU and the blue ones for $\lambda$=1 AU. 
Here one can see that for the well connected observer (the location indicated as [0,0]) decreasing the mean free path produces a strong qualitative change in the intensity profile. The time of peak intensity is delayed and the slope of the decay phase becomes less steep, as is well known from 1D transport modelling. For the not well connected observers, the change in $\lambda$ has a less pronounced effect on the overall shape of the profile and the main difference observed is the fact that the peak intensity is larger for $\lambda$=0.1 AU, since particles remain close to the Sun for a longer time and have more time to drift across the field. The slope of the decay phase varies with $\lambda$ less than for the well-connected case, and the time of peak intensity and start time of the event for $\lambda$=1 AU are in some cases later than for  $\lambda$=0.1 AU, the opposite of what would be predicted by a 1D transport model. This is because in a 3D model, at locations other than [0,0], the time variation of intensities results from the combination of drift, corotation, deceleration and scattering along the field line.
The O intensity profiles, not shown here, have a dependence on the value of the mean free path similar to that shown in Figure \ref{fig.varylambda}.

In Figure \ref{fig.ratio_lambda0.1} we present a plot of the Fe/O ratio for $\lambda$=0.1 AU: here one can see that the qualitative trend of decreasing Fe/O, for observers not well connected in the initial phase of the event, is very similar to that in Figure  \ref{fig.ratio1}, where $\lambda$=1 AU. This shows that within our 3D model the {\bf early} temporal evolution of Fe/O at observers away from the well connected field line is not strongly affected by the scattering conditions, because it is dominated by drift effects.
The increase in Fe/O seen late in the event in several cases for  $\lambda$=1 AU is no longer present for some of the observer locations for  $\lambda$=0.1 AU, due to the fact that both O and Fe fill a wider longitudinal region.

\section{Discussion}

In Section \ref{sec.test_simul} we analysed the 3D propagation of SEP Fe and O ions through interplanetary space by means of a full orbit test particle model that naturally describes the effects of drifts on heavy ions.  Fe and O were injected from a compact region at the Sun and propagated through a Parker spiral magnetic field in the presence of a low level of scattering. The ions were injected with charge states $Q_{Fe}$=15 and $Q_{O}$=7, resulting in significantly different mass-to-charge ratios. 

Our results show that, for the same energy per nucleon range,  Fe ions experience more drift than O. Both species are able to reach an observer not directly connected to the injection region, but Fe arrives and peaks first. 

Consequently, the Fe/O ratio from our simulations decays over time, in a way that qualitatively matches the behaviour observed in SEP events (see e.g. \cite{Zel2015}).
While Figures \ref{fig.intens1} and \ref{fig.ratio1} focussed on the energy range 10--30 MeV nucleon$^{-1}$, the same qualitative behaviour is observed in all SEP energy ranges within our simulation.
We also showed that the chosen value of the scattering mean free path in the simulations does not significantly affect the observed trends, since drift is only weakly dependent on the scattering conditions \citep{Mar2013}.

We conclude that propagation effects caused by drift are a possible cause of the observed temporal behaviour of the Fe/O ratio. Within the drift scenario, differences in the profiles of Fe and O are due to the 3D transport of these ions across the magnetic field, while in 1D models incorporating a rigidity dependent mean free path, they are caused by differences in the amount of scattering experienced while propagating along the field lines \citep{Sch1978,Mas2012}.
Overall in SEP events, both rigidity dependent mean free paths and drift effects may combine to produce the observed 
decays of Fe/O, though our simulations show that for an observer not well connected to the particle source the drift effects are dominant.

The intensity profiles shown in Figure \ref{fig.intens1} are shaped not only by the drift-dominated transport across the magnetic field, but also by deceleration taking place in the interplanetary medium. As discussed by  \cite{Dal2015}, drift-induced deceleration is present alongside adiabatic deceleration.

\cite{Mas2006} presented SEP observations showing that, while Fe and O intensity profiles in the same energy per nucleon range are rather different from each other, resulting in decaying Fe/O,
the profiles become almost indistinguishable when O data with average energy per nucleon double that of the Fe channel are used in the comparison.
In the latter case, the profile of Fe/O becomes flat.
The authors interpreted this behaviour as resulting from the rigidity dependence of the mean free path along the magnetic field.

The observation could however also be explained as resulting from drift processes: 
drift velocities are proportional to the product $m_0 \gamma v^2 /q$ \citep{Dal2013}, where $m_0$ is the rest mass, $\gamma$ the relativistic factor and $v$ the particle speed.
For non-relativistic particles, drift velocities are proportional to $A E /Q$ where $E$ is kinetic energy per nucleon. 
Therefore for O ions, having lower $A/Q$ than Fe, a larger value of $E$ is required for the drift velocity to be comparable to that of Fe, so as to reach a not well connected observer in similar times. 
It should be noted that drift velocities in the Parker spiral field have a dependence on position within the heliosphere \citep{Dal2013}, and particles of different energies propagate differently  e.g.~to a fixed radial distance from the Sun.
Therefore it is not possible to immediately calculate the value of the energy per nucleon of O that will result in similar drift-dominated transport to an observer, and this value is not simply related to  $(A/Q)_{Fe}/(A/Q)_{O}$ (parameter that in our simulations is 1.6).  

To investigate whether our simulations support the above qualitative explanation and compare with the observations of \cite{Mas2006}, in Figure \ref{fig.intens1_double_energy}  we consider Fe profiles for the range 10--30 MeV nucleon$^{-1}$ and O profiles in the range 30--50 nucleon$^{-1}$, at double the average energy. While there are some differences in the absolute values of the intensities, one can observe, by comparison with  Figure \ref{fig.intens1}, that the difference in arrival time between Fe and O is no longer present. 
When the corresponding Fe/O ratio is calculated, as shown in Figure \ref{fig.ratio1_double_energy}, a decay in Fe/O is no longer visible, as was the case in the observations of \cite{Mas2006}. Whether a completely flat profile is seen or not, is dependent on the location of the observer.


\section{Conclusions}

Our simulations of SEP Fe and O propagation within a Parker spiral magnetic field in the presence of weak rigidity-independent scattering have shown the following:
\begin{itemize}
\item
Significant drift is experienced by Fe and O ions away from the flux tube in which they were initially injected.
\item 
Drift-associated propagation and deceleration result in a decay over time of the Fe/O ratio at a not well connected observer, in typical SEP energy ranges.
\item
The observation that the Fe/O decay is no longer present when O at double the average energy than that of Fe is considered, is reproduced by our simulations, showing that drift alone is sufficient to explain the effect.  
\end{itemize}

We conclude that drift effects causing significant propagation across the magnetic field can qualitatively explain the observations of decaying Fe/O \citep{Sch1978,Mas2006,Mas2012,Zel2015} and the disappearance of the decaying behaviour when higher energy O is considered \citep{Mas2006}.
Therefore here we propose drift as a possible new mechanism that accounts for the observed features of Fe/O over time, alternative to the current models that explain them in terms of 1D rigidity-dependent propagation.

Our model contains a number of simplifications, which will need to be relaxed in future work to obtain a more realistic representation of SEP events. It will be necessary to move away from the simple unipolar IMF configuration and include two opposite polarities separated by a wavy heliospheric current sheet. In addition a model of field line meandering will need to be introduced  \citep{Lai2015}, which most likely will have the effect of enlarging the range of heliolongitudes and heliolatitudes over which significant intensities are detected and ensuring earlier arrival times.

A simplification introduced in our simulations is the choice of a single charge state for each of the heavy ion species considered. We judged that this assumption allows to visualise the qualitative behaviour of the Fe/O ratio in the clearest way.
In reality it is likely that Fe will be injected in the interplanetary medium with a range of charge states, although there is little observational information available on the charge profile at the acceleration site.
In a related paper \citep{Dal2016a} we analysed the propagation of Fe ions injected with a range of charge states and demonstrated that drift processes result in an energy distribution of charge states at 1 AU that increases with energy, as observed in many SEP events.
Therefore a single mechanism, drift, is able to explain both the time decay in Fe/O and the energy dependence of charge states at 1 AU, two key features of heavy ion SEP observations.

\begin{acknowledgements}
This work has received funding from the UK Science and Technology Facilities Council (STFC) (grant ST/M00760X/1) and the Leverhulme Trust (grant RPG-2015-094).
SD and TL acknowledge support from ISSI through funding for the International Team on \lq Superdiffusive transport in space plasmas and its influence
on energetic particle acceleration and propagation\rq. 
\end{acknowledgements}

\bibliographystyle{aa} 
\bibliography{heavies_biblio} 

\clearpage

\end{document}